\documentclass[aps,prl,twocolumn,groupedaddress]{revtex4}

\newif\ifpdf
\ifx\pdfoutput\undefined
\pdffalse 
\else
\pdfoutput=1 
\pdftrue
\fi

\ifpdf
\usepackage[pdftex]{graphicx}
\else
\usepackage{graphicx}
\fi
\usepackage{hyperref}

\begin{document}

\ifpdf
\DeclareGraphicsExtensions{ .pdf, .jpg}
\else
\DeclareGraphicsExtensions{.eps, .jpg}
\fi

\def\hslash{\hbar}
\def\imag{i}
\def\grad{\vec{\nabla}}
\def\div{\vec{\nabla}\cdot}
\def\curl{\vec{\nabla}\times}
\def\DDt{\frac{d}{dt}}
\def\ddt{\frac{\partial}{\partial t}}
\def\ddx{\frac{\partial}{\partial x}}
\def\ddy{\frac{\partial}{\partial y}}
\def\lap{\nabla^{2}}
\def\divv{\vec{\nabla}\cdot\vec{v}}
\def\gradS{\vec{\nabla}S}
\def\vvec{\vec{v}}
\def\wc{\omega_{c}}
\def\<{\langle}
\def\>{\rangle}
\def\Tr{{\rm Tr}}
\def\Csch{{\rm csch}}
\def\Coth{{\rm coth}}
\def\Tanh{{\rm tanh}}
\def\g2{g^{(2)}}


\title{Excited state dynamics in DNA double helices}
\author{Eric R. Bittner}
\email[email:]{bittner@uh.edu}
\homepage[URL:]{http://k2.chem.uh.edu}
\affiliation{Department of Chemistry and the Texas Center for Superconductivity, 
University of Houston, Houston TX 77204}
\date{\today\, Available as:  cond-mat/0606333}

\begin{abstract}
Recent ultrafast experiments have implicated intrachain
base-stacking rather than base-pairing as the crucial 
factor in determining the fate and transport
of photoexcited species in DNA chains. An important issue that has
emerged concerns whether or not a Frenkel excitons is sufficient 
one needs charge-transfer states to fully account for the dynamics. 
In  we present an $SU(2)\otimes
SU(2)$ lattice model which incorporates both intrachain and interchain
electronic interactions to study the quantum mechanical evolution of
an initial excitonic state placed on either the adenosine or thymidine
side of a model B DNA poly(dA).poly(dT) duplex.  Our calculations
indicate that over several hundred femtoseconds, the adenosine exciton
remains a cohesive excitonic wave packet on the adenosine side of the
chain where as the thymidine exciton rapidly decomposes into mobile
electron/hole pairs along the thymidine side of the chain.  In both
cases, the very little transfer to the other chain is seen over the
time-scale of our calculations.  We attribute the difference in these
dynamics to the roughly 4:1 ratio of hole vs. electron mobility along
the thymidine chain.
\end{abstract}


\maketitle


\begin{figure}[t]
\includegraphics[width=\columnwidth]{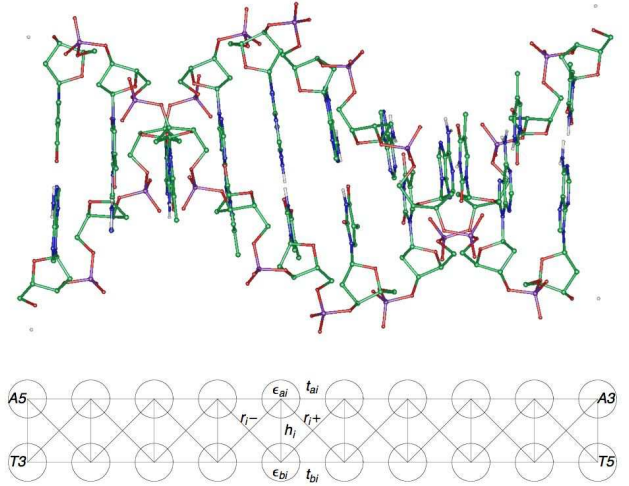}
\caption{Three dimensional structure of poly(dA)poly(dT) in the B DNA
double helical form.  This is considered to be the average structure
in water.  Bottom: Equivalent lattice model showing the connectivity
and associated transfer terms.}\label{fig1}
\end{figure}

Give the importance of DNA in biological system and its emerging role
as a scaffold and conduit for electronic transport in molecular
electronic devices, \cite{Kelley:1999} DNA in its many forms is a well
studied and well characterized system.  What remains poorly
understood, however, is the role that base-pairing and base-stacking plays in
the transport and migration of the initial excitation along the double
helix.\cite{Crespo-Hernandez:2005,Markovitsi:2005}  Such
factors are important since the UV absorption of DNA largely
represents the weighted sum of the absorption spectra of it
constituent bases whereas the distribution of lesions formed as the
result of photoexcitation are generally not uniformly distributed
along the chain itself and depend strongly upon sequence, suggesting
some degree of coupling between bases.\cite{Markovitsi:2005}

Recent work by various groups has underscored the different roles 
that base-stacking and base-pairing
play in mediating the fate of an electronic excitation in DNA.
\cite{Markovitsi:2005,Crespo-Hernandez:2005} Over 40 years ago,
L{\"o}wdin discussed proton tunneling between bases as a excited state
deactivation mechanism in DNA\cite{Lowin:1963} and evidence of this was
recently reported by Schultz {\em et al.} \cite{Schultz:2004} In contrast,
however,ultrafast fluorescence of double helix poly(dA).poly(dT)
oligomers by Crespo-Hernandez et al.\cite{Crespo-Hernandez:2005} and
by Markovitsi {\em et al.} \cite{Markovitsi:2005} give compelling
evidence that base-stacking rather than base-pairing largely
determines the fate of an excited state in DNA chains composed of A
and T bases with long-lived intrastrand states forming when ever A is
stacked with itself or with T.  However, there is considerable debate
regarding whether or not the dynamics can be explained via purely
Frenkel exciton models
~\cite{Emanuele:2005a,Emanuele:2005,Markovitsi:2006} or whether
charge-transfer states play an intermediate
role. \cite{Crespo-Hernandez:2006}

Here we report on a series of quantum dynamical calculations that
explore the fate of a localized exciton placed on either the A side or
T side of the B DNA duplex poly(dA)$_{10}$.poly(dT)$_{10}$.  Our
theoretical model is based upon a $SU(2)\otimes SU(2)$ lattice model
consisting of localized hopping interactions for electrons and holes
between adjacent base pairs along each strand ($t_{aj}$) as well as
cross-strand terms linking paired bases ($h_i$) and ``diagonal'' terms
which account for the $\pi$ stacking interaction between base $j$ on
one chain and base $j\pm 1$ on the other chain ($r^\pm_i$) in which
$r^-_j$ denotes coupling in the 5'-5' direction and $r^+_j$ coupling
in the 3'-3' direction.  Fig. ~\ref{fig1} shows the three-dimensional
structure of poly(dA)$_{10}$.poly(dT)$_{10}$ and the topology of the
equivalent lattice model.  Taking link in this figure as a
specific electron or hole hopping term,  we arrive at the following 
single particle Hamiltonian,
\begin{eqnarray}
h_{1} = \sum_j \epsilon_j \hat{\psi}_j^\dagger\hat\psi_j
+ t_j (\hat\psi_{j+1}^\dagger\hat\psi_j + \hat\psi_{j}^\dagger\hat\psi_{j+1}) )
+ h_j\overline{\psi}_j\hat\psi_j \nonumber \\
+\hat\psi_{j+1}^\dagger(r_j^+\hat\gamma_+ + r_j^-\hat\gamma_-)\hat\psi_j
+\hat\psi_{j}^\dagger(r_j^+\hat\gamma_+ + r_j^-\hat\gamma_-)\hat\psi_{j+1},
\end{eqnarray}
where $\hat\psi_j^\dagger$ and $\hat\psi_j$ are $SU(2)$ spinors that
act on the ground-state to create and remove an electron (or hole) on
the $j$th adenosine or thymidine base along the chain.  The
$\hat\gamma$ operators are the $2\times 2$ Pauli spin matrices with
$\overline{\psi}_j = \hat\gamma_1\hat\psi_j^\dagger$ and
$\hat\gamma_++\hat\gamma_- = \hat\gamma_1$ providing the mixing
between the two chains.  Taking the chain to homogeneous and infinite
in extent, one can easily determine the energy spectrum of the valence
and conduction bands by diagonalizing
\begin{eqnarray}
\hat{h}_{1}=
\left(
\begin{array}{cc}
\epsilon_{a} + 2 t_{a} \cos(q)   &  h + r^{+}e^{-iq} + r^{-}e^{+iq} \\
h + r^{+}e^{+iq} + r^{-}e^{-iq} \    & \epsilon_{b} + 2t_{b} \cos(q)
\end{array}
\right)\label{bandop}
\end{eqnarray}
where $\epsilon_{a,b}$ and $t_{a,b}$ are the valence band or conduction band site energies 
and intra-strand hopping integrals.
When  $r_j^+ = r_j^-$,  Eq.~\ref{bandop} is identical to the Hamiltonian 
used by Creutz and Horvath~\cite{Creutz:1994} to describe chiral symmetry in quantum 
chromodynamics in which  the terms proportional to $r$ are introduced to 
make the ``doublers'' at $q\propto \pi$ heavier than the states at $q \propto 0$. 

The coupling between the conduction and valence bands
is accomplished by introducing short-ranged Coulomb and exchange interactions as well as 
``dipole-dipole'' terms which couple geminate electron-hole pairs on different sites.  
\begin{eqnarray}
H(12) = h_1 + h_2 + \sum_{{\bf m,n}} V_{{\bf m,n}} A_{\bf m}^\dagger A_{\bf n}
\end{eqnarray}
where the $A_{\bf m}$ are spin-symmetrized composite operators that
create or remove singlet or triplet electron/hole pairs in
configuration $|{\bf m}\rangle = |i_e j_h\rangle$ where $V_{\bf mn} =
-\langle m_en_h||n_e m_h\rangle + 2\delta_{S0}\langle m_e n_h || n_e
m_h\rangle$ where $S = 1,0$ is the total spin.
\cite{karabunarliev:4291,karabunarliev:3988,karabunarliev:10219,karabunarliev:057402}

The single particle parameters are taken from Anatram and Mehrez as
determined by computing the Coulomb integrals between HOMO and LUMO
levels on adjacent base pairs with in a double-strand B DNA sequence
using density functional theory (B3LYP/6-31G)~\cite{Mehrez:2005}.  For
(dA).(dT), the intrachain electron and hole transfer terms are -0.023
eV and -0.098 eV for the thymidine chain and +0.024 and +0.021 eV
along the adenosine chain.  The interchain terms are $h_j = 0.063$ eV,
$r_j^+ = -0.012$eV, and $r_j^-=-0.016$eV for the electron and
$h_j=0.002$eV, $r_j^+=-0.007$eV, and $r_j^-=0.050$eV for the hole with
site energies $\varepsilon_e = -0.931$eV and $\varepsilon_h =
-6.298$eV for the thymidine chain and $\varepsilon_e = 0.259$eV and
$\varepsilon_h = -5.45$eV for the adenosine chain.  It is important to
note that the asymmetry introduced with $r_j^+\ne r_j^-$ gives
directionality between the 3'- and 5'- ends of the chain.

For simplicity, we take the on-site Coulomb interaction $J = -2.5$ eV
and the on-site exchange interaction to be $K = 1.0$eV for both the
purines and pyrimidines.  We assume these interactions to be local
since the distance at which the Coulomb energy between an
electron/hole pair equals the thermal energy in aqueous ionic media at
300K is on the order of the base-stacking distance.  These we set as
adjustable parameters to tune the predicted absorption spectrum of our
model to reproduce the experimental UV absorption
spectra.~\cite{Bittner:2006} Lastly, we estimated the coupling between
geminate electron/hole pairs on different bases $\langle n_e n_h ||
m_e m_h\rangle$ via a point-dipole approximation by mapping the
$\pi-\pi^*$ transition moments onto the corresponding base in the B
DNA chain.  \footnote{These are obtained from the isolated bases by
performing single configuration interaction (CIS) calculations using
the GAMESS\cite{GAMESS} quantum chemistry package on the corresponding
9-methylated purines and 1-methylated pyrimidines after optimizing the
geometry at the HF/6-31(d)G level of theory.}  As in the electron and
hole hopping terms, the local dipole-dipole coupling terms read:
$d_\perp = -0.099$ eV between adjacent base pairs, $d^A_\parallel =
0.0698$eV along the adenosine chain, $d^T_\parallel = 0.143$eV along
the thymidine, $d^- = -0.006$eV for the 5'-5' diagonal coupling and
$d^+ = -0.013$eV for the 3'-3' diagonal coupling.  The use of the
point-dipole approximation in this case is justified mostly for
convenience and given the close proximity of the bases, multipole
terms should be included in a more complete model.
\cite{Bouvier2002vu} As a result, the matrix elements used herein
provide an upper limit (in magnitude) of the couplings between
geminate electron-hole pairs.  Most importantly, however, the
point-dipole approximation provides a robust means of incorporating
the geometric arrangement of the bases into our
model.\cite{Bittner:2006} Having briefly discussed the model and how
we determined the parameters, we move onto discuss the result of our
dynamical calculations.

We consider the fate of an initial singlet electron/hole pair placed
either in the middle of the thymidine side of the chain or the
adenosine side of the chain (i.e. an exciton).  We assume that such a
configuration is the result of an photoexcitation at the appropriate
photon energy (4.87 eV for the thymidine exciton and 5.21 eV for the
adenosine exciton respecitively) and based upon the observation that
the UV absorption spectra largely represents the weighted sum of the
UV spectra of the constituent bases, such localized initial states are
Since these are not stationary states, they evolve according to the
time-dependent Schr{\"o}dinger equation. Over a short time-step,
$\delta t$ this is easily computed using the Tchebychev expansion of
the time-evolution operator~\cite{Tal-Ezar:1981}
\begin{eqnarray}
\Psi(t+ \delta t) &=& \exp\left[-i H(12)\delta t /\hbar\right]\Psi(t) \nonumber \\&=&  
e^{-i\overline{E}\delta t/\hbar}\sum_{n=0}^{M}a_{n}(\alpha)T_{n}(-i\tilde H)|\Psi(t)
\end{eqnarray}
$\Delta E = E_{max}- E_{min}$.
where  $T_{n}(x)$ are Chebychev polynomials. 
\begin{eqnarray}
T_{n+1}(x) = 2 x T_{n}(x) - T_{n-1}(x)
\end{eqnarray}
with $T_{0}(x) = 1$ and $T_{1}(x) = x$.  
$\overline H$ is Hamiltonian operator shifted and scaled so that its eigenvalues lie within
$E \in [-1,1]$, and  $J_{n}(\alpha)$ the $n$th spherical Bessel function with 
$\alpha = \Delta E t/2\hbar$. 
The advantage of this approach is that converges rapidly, preserves 
norm, and is computationally 
efficient since it involves at most matrix-vector multiplications.

\begin{figure}[t]
\includegraphics[width=\columnwidth]{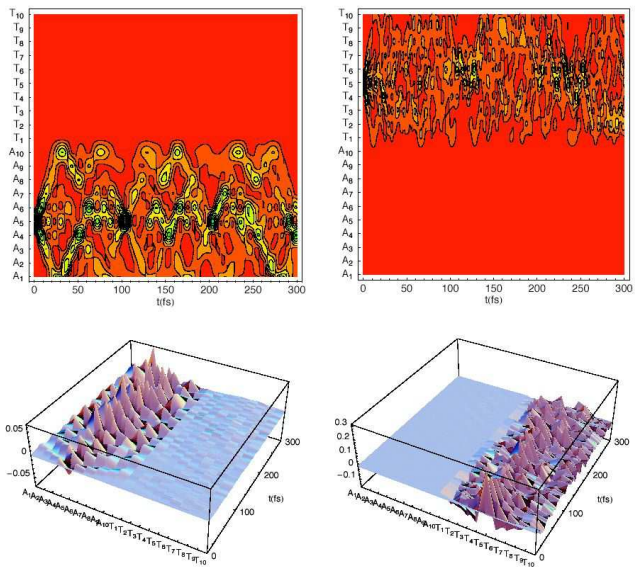}
\caption{Top: Time evolution of the exciton density for an initial
excitation placed on the adenosine (left) and thymidine (right)
chains).  Bottom: The corresponding net charge on a given site
following excitation of the adenosine (left) and thymidine (right)
chains.}
\label{fig2}
\end{figure}

In the top two frames of Fig.~\ref{fig2} we show the transient
probability for finding an exciton placed on the adenosine (left) or
thymidine (right) chain at time $t=0$ in some other excitonic
configuration along either the adenosine or thymidine chain at some
time $t$ later.  In both cases, negligible exciton density is
transferred between chains and the excitons rapidly become delocalized
and scatter ballistically down the DNA chain.

There are some striking differences, however, between the exciton
dynamics in adenosine versus those in thymidine.  First, in comparing
the $d_\parallel$ matrix elements, one easily concludes that the
exciton mobility along the thymidine chain is considerably greater
than the mobility along the adenosine chain.  This can be see in Fig
2. comparing the time required for an excitonic wavepacket to reach
the end of either chain.  In adenosine, the exciton travels 5 base
pairs in about 25 fs where as an exciton along the thymidine chain
covers the same distance in about 10 fs.  This factor of two
difference in the exciton velocity is commensurate with the
$\approx$1:2 ratio of the $d_\perp^A:d_\perp^T$ intrachain excitonic
couplings.

Secondly, we note that the adenosine exciton remains qualitatively
more ``cohesive'' than the thymidine exciton showing a number of
ballistic traverses up and down the adenosine chain over the 300 fs we
performed the calculation.  One can also note that the exciton
velocity in the 5'-3' direction is slightly greater than in the 3-5'
direction as evidenced by the exciton rebounds off site $A_1$ slightly
sooner than it rebounds from site $A_{10}$.  This is due to the
asymmetry introduced in by the $r^\pm$ and $d^\pm$ terms.  All in all,
one can clearly note a series of strong recurrences for finding the
adenosine exciton on the original site $A_5$ every 100 fs.  The
thymidine exciton dynamics are far more complex as the exciton rapidly
breaks apart.  While few recursions can be noted, however, after the
first ballistic traverse, the thymidine exciton no longer exists as a
cohesive wavepacket and is more or less uniformly distributed along
the thymidine side of the chain.

The excitonic dynamics only tell part of story.  In the lower two
frames of Fig. \ref{fig2} we show the net charge taken as the
difference between the hole density and electron density on a given
base.  Note the difference in scale between the bottom left and bottom
right figures.  In the case where the initial exciton is on the
adenosine chain, very little charge-separation occurs over the time
scale of our calculation.  On the other hand, when the exciton is
placed on the thymidine chain, the exciton almost {\em immediately}
evolves into a linear combination of excitonic and charge-separated
configurations.  What is also striking is that in neither case do
either the electron or hole transfer over to the other chain even
though energetically charge-separated states with the electron on the
thymidine and the hole on the adenosine sides of the chain are the
lowest energy states of our model.~\cite{Bittner:2006} It is possible,
that by including dissipation or decoherence into our dynamics, such
relaxation will occur, however, on a time scale dictated by
cross-chain transfer terms.  For the coupling terms at hand, electron
or hole transfer across base pairs occurs on the time scale of 3-4 ps.

\begin{figure}[b]
\includegraphics[width=\columnwidth]{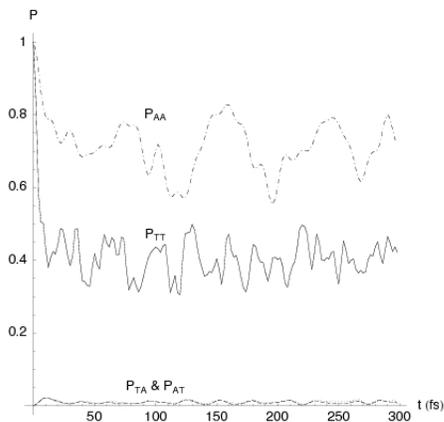}
\caption{Net probability, $P_{if}$ for the initial exciton to remain an exciton on the 
initial chain $(P_{AA} \& P_{TT})$ or to be transferred as an exciton the 
other chain $(P_{AT} \& P_{TA})$}\label{fig3}
\end{figure}

The difference between the excitonic dynamics following excitation of
A vs. T can be quantitatively noted by comparing the curves shown in
Fig.~\ref{fig3} where we compare the projection of the time-evolved
state onto the excitonic configurations of the chain on which the
exciton was placed ($P_{AA}$ and $P_{TT}$) compared to the projection
onto the excitonic configurations of the other chain ($P_{AT}$ and
$P_{TA}$).  For the case in which the adenosine chain was excited,
approximately 75\% of the total probability density remains as
excitonic configurations along the adenosine chain.  In stark
contrast, only about 40\% of the initial thymidine exciton remains
excitonic along the thymidine side of the chain.  The reason for the
remarkable difference between the two chains stems from the difference
in electron and hole mobility along the thymidine chain.  Indeed,
comparing the electron and hole hopping terms given above,
$t_{h}/t_{e} \approx 4$ for the thymidine chain compared to
$t_{h}/t_{e} \approx 1$ for along the adenosine chain.  This is
manifest in the lower right panel of Fig. 2 where we see almost
immediately a negative charge remaining for a few fs on the site where
the initial excitation was placed.

The results described herein paint a similar picture to that described
by recent ultrafast spectroscopic investigations of (dA).(dT)
oligomers in that the initial excitonic dynamics is dominated by
base-stacking type interactions rather than by inter-base couplings.
Interchain transfer is multiple orders of magnitude slower than the
intrachain transport of both geminate electron/hole pairs as excitons
and independent charge-separated species.
Indeed, for an exciton placed on the adenosine chain, our model
predicts that exciton remains as a largely cohesive and geminate
electron/hole pair wave function as it scatters along the adenosine
side of the chain.  Our model also highlights how the difference
between the mobilities in the conduction and valence bands localized
along each chain impact the excitonic dynamics by facilitating the
break up of the thymidine exciton into separate mobile
charge-carriers.  In the actual physical system, the mobility of the
free electron and hole along the chain will certainly be dressed by
the polarization of the medium and reorganization of the lattice such
that the coherent transport depicted here will be replaced by
incoherent hopping between bases.

In conclusion, we present herein a rather compelling model for the
short-time dynamics of the excited states in DNA chains that
incorporates both charge-transfer and excitonic transfer.  It is
certainly not a complete model and parametric refinements are
warranted before quantitative predictions can be established.  For
certain, there are various potentially important contributions we have
left out: disorder in the system, the fluctuations and vibrations of
the lattice, polarization of the media, dissipation, decoherence, etc.
These we recognize as lacunae in our model.  


{\em Acknowledgments:} This work was funded by the National Science
Foundation, the Robert Welch Foundation, and the Texas Center for
Superconducivity.  The author also wishes to thank Prof. Stephen
Bradforth (U. So. Cal.) for many insightful conversations leading to
this work.


\end{document}